\documentstyle[preprint,aps,eqsecnum]{revtex}
 
\begin{document}
\draft
 
\title{Dissipative dynamics of vortex lines in superfluid $^{4}$He}
 
\author{H. M. Cataldo, M. A. Desp\'osito, E. S. Hern\'andez and D. M.
Jezek}
\address{
Departamento de F\'{\i}sica, Facultad de Ciencias Exactas y Naturales, \\
Universidad de Buenos Aires, RA-1428 Buenos Aires, Argentina\\
and Consejo Nacional de Investigaciones Cient\'{\i}ficas y
T\'ecnicas, Argentina}

\date{\today}
\maketitle
 
\begin{abstract}

	We propose a Hamiltonian model that describes the interaction
between a vortex line in superfluid $^{4}$He and the gas of
elementary excitations. An equation
of irreversible motion for the density operator of the vortex,
regarded as a macroscopic quantum particle with a finite mass,
is derived in the frame of Generalized Master Equations. This
enables us to cast the effect of the coupling as a drag force
with one reactive and one dissipative component,  in agreement
with the assumption of the phenomenological theories of vortex
mutual friction in the two fluid model.
\end{abstract}

 
\narrowtext

\section{Introduction}

	Since the discovery of quantized vortices in liquid helium
II, it has been recognized that they might provide a mechanism for
the coupling of the superfluid to the normal fluid. In the two fluid
model, this coupling is represented by a mutual drag force with one
dissipative and one conservative component, whose respective
strengths can be measured investigating the attenuation of second
sound at various temperatures \cite{HV,Don1,Don2}. Models for
the friction coefficients which successfully fit the data up to
 2.1 K have been presented in Refs. \cite{MS,SD}.
The phenomenon of vortex mutual friction has been observed as
well in rotating superfluid $^{3}$He-B \cite{Bev}; this fact
brings support to the conjecture that such a mechanism is indeed
a relevant source of dissipative processes in highly degenerate
quantum fluids.   It is also worthwhile to remind that in the
last years, vortices  have been seen to play a role
in phase transitions taking place either in underpressurized
$^{4}$He \cite{Ma} or in supersaturated solutions of $^{3}$He in
$^{4}$He \cite{DJ}.

 The interaction between the velocity field and the
density fluctuations of the superfluid,  which at nonvanishing
temperatures are embodied in the normal fluid, can be accounted
for within a Lagrangian description \cite{dem1,dem2}. However,
microscopic descriptions of the interaction between the
superfluid motion, especially when topological singularities are
concerned, and collective excitations, are not conclusive
\cite{Don1,Don2}. The investigation of quantum tunneling of
vortex lines in superconductors and superfluids \cite{Ao,Ste,Niu}
has now  improved our comprehension of, for example, the role
of the inertial mass of the vortex (see also Refs.
\cite{Duan0,Duan1,Duan2} and cited therein) and the influence of
either pinning or dissipation on the tunneling rates. Indeed,
the value of the vortex inertia is a fundamental parameter in
any theoretical description of vortex dynamics and it remains being a
controversial issue \cite{Duan2,Niu3}. Different starting
points assign to this inertia figures ranging from zero to infinity, the
latter arising from a logarithmic divergence with the system
size due to the renormalization effect induced by the condensate
motion.  On the other hand, the vortex mass is known to be
finite in superconductors \cite{Su}. Although the
phenomenological approaches
\cite{Don2} completely disregard inertial effects,
 the mass enters the description of the dynamics of a free
vortex, known to be cyclotron-like \cite{Don1,dem2,Niu3}, 
through a frequency parameter $\Omega$, which would be a
measurable quantity if this time dependent regime were
experimentally visualized. In particular, it has been recently
shown \cite{dem2} that the cyclotron motion is a natural
solution of the nonlinear Schr\"{o}dinger equation applied to
a vortex. Dynamical \cite{dem2} and thermodynamical \cite{Mo}
methods have been proposed to measure the vortex inertial
coefficient; our present purpose is neither to participate in the
existing polemics, nor to propose a new model for the
calculation of the vortex mass, but rather to assume that it is
a  numerical parameter and proceed along similar lines as those
invoked in the well established cyclotron motion already
discussed in textbooks \cite{Don1}.

	The aim of the present work is to propose a
Hamiltonian model for the coupling between a rectilinear
vortex immersed into the  excitations of the superfluid, as will be 
discussed in Sec. 2. Due to the translational symmetry of the
problem along an axis parallel to the vortex, the problem is a
twodimensional one, {\it i. e.}, we consider a point vortex on a
plane. We shall show in Sec. 3 that, if one  considers the vortex as a
quantum particle undergoing  Brownian motion \cite{DH} in
a heat reservoir, it is possible to establish the irreversible
time evolution of its density operator within the Generalized
Master Equation (GME) approach
\cite{Haa,Cha}.  In this way, in Sec. 4 we are able to derive dissipative
equations of motion for the canonical position-momentum
variables of the vortex and for its velocity.
These variables  can be seen to evolve under the combined effect
of the usual hydrodynamical lift on a rotating cylinder, plus a
drag force. If the coupling is linear in the excitation
operators, the drag  coefficients are governed by the dynamical
susceptibility of the liquid. The consequences of the equations
of motion thus obtained, the asymptotic velocity of the vortex
and the relation of the current description to the phenomenological model
are discussed in Sec. 5, where the perspectives of the present
approach are also outlined.

\section{The Hamiltonian model}

Let us first summarize the description of the free motion.
 The  Hamiltonian  for 
 a cylindrical vortex parallel to the $z-$axis in liquid
helium at zero temperature is in charge of providing the Magnus
force \cite{Don1}.  It reads
\begin{equation}
H_v = \frac{1}{2 M} \left [{\bf p} - q {\bf A}({\bf r})\right]^2
+ M \, \Omega\, v_s\, y
\label{Hfree}
\end{equation}
where
\begin{equation}
{\bf A}({\bf r}) = \frac{h\,\rho_s\,l}{2} (y,-x)
\end{equation}
is the vector potential whose curl yields the vortex-velocity-dependent
part of the Magnus force and the potential term $M \Omega v_s y$
 gives the superfluid-velocity-dependent part of this
force. Here $M$ is the
dynamical mass of the vortex, $\rho_s$ denotes the number
 density of the superfluid, $v_s$ its velocity along the $x-$axis, assumed
to be uniform,  $h$ is Planck's constant, $l$ the system length
along the $z-$ axis and
\begin{equation}
\Omega= \frac{q \,h\,\rho_s\,l}{M}.
\end{equation}
 The quantity $q = \pm 1$
is the sign of the vorticity according to the right handed
convention. Furthermore, at zero temperature, 
 $\rho_s$ coincides with the total  density per unit mass
$\rho/m$, being $m$ the mass of a helium atom.

 At this point it is convenient to remember the existing
theoretical uncertainty regarding the vortex mass parameter $M$
that should appear in dynamical calculations
\cite{Niu,Duan1,Duan2} and keep in mind that in the
phenomenological approaches \cite{Don2}  the
dynamical regime of the vortex is that in which the Magnus force
balances the drag plus any applied force \cite{BDV}. As stated
in the Introduction, our viewpoint here is identical to that of
former authors \cite{Don1,dem2,Ao,Niu} who assume a finite
figure for the vortex inertia and consider the cyclotron - like
motion of a free vortex as their starting point, with the
frequency $\Omega$ as the leading parameter. Since it will be  shown
in Sec. 4 that the dissipative motion is easily described in
terms of the complex position variable $z = x + i y$ and the
velocity $d z/d t$, we here write the complex Hamilton equation
that stems from (2.1)
\begin{equation}
\frac{ d^2 z}{d t^2} = i\,\Omega\,\left(\frac{d z}{d t} - v_s\right)
\end{equation}
with the complex Magnus force at the right hand side.

We now assume liquid helium to contain elementary excitations.
For nonvanishing temperatures $T$, these excitations can be of
thermal origin and thus give rise to the normal fluid, while at
zero temperature they must be created by an external probe and
yield a vanishing normal density $\rho_n$.  If $T$ is above 1 K,
the normal component is mainly a gas of rotons, being the
phonons the dominant excitations at lower temperatures.
Therefore, at any temperature  the interaction of the vortex
line with these elementary excitations, produces  damped
motion of the vortex.  As stated in the introduction, the
main goal of this article is to construct a hamiltonian model
which enables us to obtain this dissipative behavior.

 For this purpose, we shall
consider a  description of quantum dissipation similar to 
that recently presented in order to account for the
irreversible evolution of solitons \cite{CNCal2}, in which an effective
Hamiltonian is constructed for the collective motion coupled to the
residual excitations using the Collective Coordinate formalism
 \cite{Raj}. This model exhibits unexpected
features \cite{MD} due to the fact that both
the system and the reservoir  have the same microscopic origin,
which is just the case here discussed.

In this spirit, and considering that within a superfluid in its
ground state the vortex exhibits a soliton-like behavior, we
propose a vortex-plus-reservoir Hamiltonian, that 
modifies expression (\ref{Hfree}) as follows
\begin{equation}
H = \frac{1}{2 M} \left [{\bf p} - q {\bf A}({\bf r}) - \lambda\,{\bf
B}\right]^2 + M\,\Omega\,v_s\,y + H_B
\label{H}
\end{equation}
with ${\bf B}$ a vector function of operators that represents the
elementary excitations of the superfluid  and $H_B$ is the
Hamiltonian of these excitations. In Eq. (\ref{H})
the interaction term $H_{int} = - \lambda\, {\bf B}\cdot
{\bf v}$ couples the  reservoir and the vortex  through the
unperturbed velocity of the latter, being
\begin{equation}
{\bf v} = \left(\frac{ p_x}{M}- \frac{\Omega}{2} \,y, \frac{ p_y}{M} +
\frac{\Omega}{2}\,x\right).
\label{vfree}
\end{equation}

In the present approach, the hermitian operator ${\bf B}$   is
associated to the creation of a density fluctuation in the liquid
 and could then be  labelled by a
transferred momentum ${\bf q}$. Up to lowest order, one may have
 for each component of the vector ${\bf B}$
\begin{equation}
B_{\bf q} = \frac{\hat{O}^{\dagger}_{\bf q} + \hat{O}_{\bf
q}}{\sqrt{2}}
\label{B}
\end{equation}
where $\hat{O}^{\dagger}_{\bf q}\, (\hat{O}_{\bf q})$ is the
Feynman-Cohen operator that creates (destroys) a density
fluctuation quantum, {\it i.e.}, a phonon or a roton
\begin{equation}
\hat{O}^{\dagger}_{\bf q} = \rho^{\dagger}_{\bf q} -\frac{1}{N}
\sum_{{\bf k} \neq {\bf q}}
\frac{{\bf k}\cdot {\bf q}} {k^2}\,\rho^{\dagger}_{\bf
k}\,\rho^{\dagger}_{{\bf q} - {\bf k}}.
\label{O}
\end{equation}
 being here N the number of atoms in the liquid.
Furthermore, we realize that the term $\lambda^2\,B^2/2 M$
appearing in (2.5) can be absorbed into the hamiltonian $H_B$,
which is in charge of providing the equilibrium density vector
of the reservoir.

\section{The Generalized Master Equation}

	The Hamiltonian (\ref{H}) is of the form
system - plus - reservoir - plus - interaction \cite{Haa}. The
standard reduction - projection procedure of nonequilibrium
statistical mechanics \cite{DH}, enriched with the time
convolutionless method developed by Chaturvedi and Shibata\cite{Cha}, has
already proven to be useful to derive a generalized master
equation (GME), with time dependent coefficients,  for the density
operator $\sigma$ of a particle interacting with a heat reservoir in
the weak coupling - nonmarkovian limit
\cite{Mad1,Mad2}. In this case, $\sigma$ is the density operator of the
vortex and the generalized master equation reads \cite{MD}
\widetext
\begin{eqnarray}
\label{emg}
\frac{d\,\sigma}{d\,t}+\frac{i}{\hbar} [H_v,\sigma] & = &
-\frac{\lambda^2}{\hbar^2}\,\int_0^t d \tau\, 
\left\{[v_x,[v_x(-\tau),\sigma]]+[v_y,[v_y(-\tau),\sigma]]\right\}
\phi(\tau)
 \nonumber \\
&& -i \frac{\lambda^2}{\hbar^2}\,\int_0^t d \tau\, 
\left\{[v_x,[v_x(-\tau),\sigma]_{+}]+[v_y,[v_y(-\tau),\sigma]_{+}]\right\}
\psi(\tau)
\end{eqnarray}
\narrowtext
\noindent where $[a,b]_{+}$ denotes an anticommutator. 
In this expression, the time dependent functions $\phi$ and $\psi$
are the real and imaginary parts, respectively, of the correlation
between heat bath operators \cite{esh},
\begin{equation}
<B_j(\tau)\,B_j> = \phi(\tau) + i \psi(\tau)
\end{equation}
for $j = x, y$, assuming an isotropic reservoir. If the
hermitian operator $B_j$ is  chosen according to Eqs.
(\ref{B}), (\ref{O}),  the function
\begin{equation}
S_{\bf q}(\tau) = \phi_{\bf q}(\tau) + i\,\psi_{\bf
q}(\tau)
\end{equation}
 is just the Fourier transform of the dynamical
structure factor $S({\bf q},\omega)$ of helium II and is
experimentally known for a wide range of transferred momenta \cite{Gri,Gly}.

Notice that the GME is a differential, rather than an
integrodifferential, equation, since the unknown $\sigma$ under the
integral sign is taken at time $t$; accordingly, it can be simplified
if we define the following time dependent parameters,
\begin{equation}
\frac{M}{2} \gamma(t) = - \frac{\lambda^2}{\hbar}\,\int_0^t d
\tau\,\psi(\tau)\,{\rm sin}
\Omega \tau
\label{gama}
\end{equation}
\begin{equation}
\frac{M}{2} \mu(t) = \frac{\lambda^2}{\hbar}\,\int_0^t d
\tau\,\psi(\tau)\,{\rm cos}
\Omega \tau
\label{mu}
\end{equation}
and
\begin{equation}
C(t) = \frac{\lambda^2}{\hbar^2}\,\int_0^t d
\tau\, \phi(\tau)\,{\rm cos}
\Omega \tau.
\label{C}
\end{equation}

The velocities appearing in Eq.  (\ref{emg}) are those of the
free vortex displayed in (\ref{vfree}) and their detailed
time dependence is extracted from 
 Hamilton's equations corresponding to the
Hamiltonian (\ref{Hfree}), namely
\begin{eqnarray}
v_x(t)& = &[v_x(0) - v_s]\,{\rm cos} \Omega t - v_y(0)\,{\rm sin}
\Omega t+ v_s
\nonumber
\\
v_y(t) & = &  [v_x(0) - v_s]\,{\rm sin} \Omega t + v_y(0) {\rm cos}
\Omega t.
\label{vt}
\end{eqnarray}
In terms of these quantities and using Eqs. (3.4) to (3.6), we can
write
\widetext
\begin{eqnarray}
\frac{d\,\sigma}{d\,t}+\frac{i}{\hbar} [H_{eff},\sigma] & = &
 -C(t)\,\left\{[v_x,[v_x,\sigma]]+[v_y,[v_y,\sigma]]\right\}
\label{em}
\\
&& +\frac{i} {\hbar}\,\frac{M \gamma(t)}{2}
\left\{[v_x,[v_y,\sigma]_{+}]-[v_y,[v_x,\sigma]_{+}]\right\}.
\nonumber
\end{eqnarray}
\narrowtext
The effective Hamiltonian contains a renormalization to the vortex
mass, induced by the coupling to the thermal reservoir, plus a drift
contribution. Its expression is
\widetext
\begin{equation}
H_{eff} = H_v + \frac{M \mu(t)}{2} \left(v_x^2+v_y^2\right) +
M\, v_s\,\omega(t)\, v_x -
M\, v_s\,\gamma(t)\, v_y
\label{heff}
\end{equation}
\narrowtext
\noindent where 
$\omega(t) = \left.\mu(t)\right|_{\Omega = 0} - \mu(t)$.
It is also worthwhile noticing that, being the system
translationally invariant on the $(x,y)$ plane, terms in
$H_{eff}$ proportional to $v_x, v_y$ play no role in the
dynamics.

It is important to observe that the validity of the GME
(\ref{em}) is more general than  the weak coupling
approximation case.  Indeed, if one expands the integral, time
dependent collisional kernel of the master equation  
 in powers of the coupling
parameter $\lambda$, as done, for instance, in Ref.
\cite{Cha},  after a lengthy calculation one
can realize that the form of the new GME is identical to (\ref{em}), at
least up to the fourth order in the expansion parameter, 
except for the fact that the coefficients (\ref{gama}) to
(\ref{C}) become polynomials in $\lambda$.

Finally, we should also mention that the most common
 assumptions considered in many applications are that  the reservoir is
purely harmonic and/or that the interaction term  is linear in its
 coordinates. If this is not the current case,
the nonvanishing mean values $<B_j>$ must be
considered \cite{MD} and modify the effective Hamiltonian
({\ref{heff}). However, the equations of motion 
that we shall derive in the next section remain invariant, since
these extra terms can be removed by a Galilean transformation.
Moreover, it should be noticed as well that the correlation
function $<B_k(\tau)\,B_j> $ for $k\ne j$, which is in general
a nonvanishing function, does not enter Eq.  (\ref{emg}).

\section{The equations of irreversible motion}

We are now in a position to derive equations of motion for
expectation values $\langle a\rangle$ of
arbitrary observables $a$, which can be cast in the form
\widetext
\begin{eqnarray}
\frac{d\,\langle a\rangle}{d\,t}+\frac{i}{\hbar}\, <[a,H_{eff}]> &=&
-C(t) \left(<[[a,v_x],v_x]> +<[[a,v_y],v_y]>\right)
\nonumber
\\
&+& \frac{i}{\hbar}\,\frac{M \,\gamma(t)}{2} 
\left(<[[a,v_x],v_y]_{+}> -<[[a,v_y],v_x]_{+}>\right).
\end{eqnarray}
\narrowtext
In order to derive equations of motion for
the position and momentum components of the vortex
 we will restrict ourselves to the markovian limit; in
other words, we consider that the correlation indicated by
$\phi,\psi$ is short lived, within the observational times.
 The parameters in Eqs.
(\ref{gama}) and (\ref{mu}) become then time independent and after
some algebra, elimination of the momentum 
 permits us to write
a unique complex differential equation  for the expectation
value of its velocity, that  exhibits the effects of the
coupling to the reservoir.  This equation is
\begin{equation}
\frac{d^2\,\langle z\rangle}{d\,t^2}= i \Omega\,\beta\,
\left(\frac{d\,\langle z\rangle}{d\,t} - v_s \right)
\label{z}
\end{equation}
where the quantity that renormalizes the complex Magnus force 
 is
\begin{equation}
\beta= 1 + \mu + i \,\gamma.
\end{equation}
with $\mu$ and $\gamma$ being the asymptotic values of (\ref{mu}) and
(\ref{gama}) respectively.

We here realize that the reservoir constituted by the excitations of
the superfluid provides both a dissipative and a conservative
coupling, respectively measured by the parameters $\gamma$ and $\mu$.
This is in agreement with the structure of the mutual friction
force of the two fluid model \cite{HV,Don1,Don2}. Moreover,
keeping in mind that if the density fluctuations of the
superfluid carry a definite momentum
${\bf q}$, the heat bath correlation $\langle B_{\bf q} B_{\bf
q}(\tau)\rangle$ is, to lowest order, just the Fourier transform
of the dynamical structure factor, using standard relations of
linear response theory \cite{pin} one can readily show that
\begin{equation}
\mu + i \gamma = \frac{\lambda^2\,l}{2 \pi \hbar M}\,\chi({\bf
q},\Omega)
\label{chi}
\end{equation}
where $\chi({\bf q},\Omega)$ is the dynamical susceptibility or
response function of the liquid (per unit length) at momentum
${\bf q}$ and energy $\hbar \Omega$. In the most general
situation where thermal excitations cover the whole momentum
range, a summation over ${\bf q}$ should be applied on the right
hand side of Eq.  (\ref{chi}). A consequence of this result is
that the temperature dependence of the drag coefficients is
provided by the variation of $\chi({\bf q}, \Omega)$ with $T$
\cite{Gly}; it is then worthwhile to keep in mind that the
harmonic oscillator heat bath employed in previous
investigations of vortex coupling to excitations
\cite{Ao,Niu} predicts a temperature independent dissipation
strength \cite{esh}. However, if the reservoir operators $B_j$
are described by nonlinear functions of $\hat{O}_{\bf
q}^{\dagger},\hat{O}_{\bf q}$ rather than by the Feynman-Cohen
operator (\ref{O}), it is possible to show that the drag
coefficients vanish at zero temperature \cite{CNCal2,MD}; in
such a case, the coupling is uneffective and the vortex moves
freely governed by the Hamiltonian (\ref{Hfree}), as expected.

Equation (\ref{z})  can be straightforwardly integrated, giving a mean
value of the complex velocity operator
\begin{equation}
\frac {d \langle z\rangle}{d t} = 
\left[\left.\frac{d \langle z\rangle}{d t}\right|_{t=0} -
v_s\right]\,e^{\displaystyle  i \Omega
(1 + \mu)t}\,e^{\displaystyle - \Omega\gamma t} + v_s.
\label{vl}
\end{equation}
Since $\Omega\,\gamma$ is always a positive quantity, this expression
contains exponential decay of the initial conditions, and  the
limiting value of the vortex velocity is then the superfluid one
$v_s$.    

Let us now examine the situation as
described by  the phenomenological theory \cite{HV,Don1,Don2}
where the drag force is written as
\begin{equation}
 f_D = - (\gamma_0+ i \gamma'_0)\,\left(\frac{d  z}{d t} - v_n\right)
\label{drag}
\end{equation}
where $\gamma_0, \gamma'_0$ are the
strenghts of the dissipative and conservative components and
$v_n$ is the normal fluid velocity. It is also assumed that when
equilibrium is reached, $f_D$ can be expressed in the form $
(\alpha - i \alpha')\,\rho_s\,q\,h\, (v_n - v_s)$. 	If one
solves Newton's equation for a point particle with mass $M$
moving under the Magnus and the drag force (\ref{drag}), one
finds
\widetext
\begin{eqnarray}
\frac {d  z}{d t} &= &
\left[\left.\frac{d  z}{d t}\right|_{t=0} -
v_s-(\alpha'+i\,\alpha)\,(v_n - v_s)\right]\,e^{\displaystyle  i (\Omega
- \gamma'_0\,l/M) t}\,e^{\displaystyle - \gamma_0\,l t/M} 
\nonumber
\\
& + & (\alpha' +
i\,\alpha)\,(v_n - v_s)+ v_s
\label{feno}
\end{eqnarray}
\narrowtext
\noindent where $\alpha$ and $\alpha'$ can be written in terms
of $\gamma_0$ and $\gamma'_0$ as 
\begin{eqnarray}
\alpha&=  &\rho_s\,q\,h\, \frac{\displaystyle \gamma_0}{\displaystyle
(\rho_s\,q\,h-\gamma'_0)^2+\gamma_0^2}
\nonumber
\\
\alpha'&=&\frac{\displaystyle \gamma_0^2 
+\gamma'_0\,(\gamma'_0 - \rho_s\,q\,h)}{\displaystyle
(\rho_s\,q\,h-\gamma'_0)^2+\gamma_0^2}
\label{rela}
\end{eqnarray}
The inverse relationships giving $\gamma_0, \gamma'_0$ in terms
of $\alpha, \alpha'$ can be found in Ref. \cite{Don1}.   We see
that in this case, the asymptotic velocity contains both the
reactive and the resistive coefficients. However, measurements
of second sound attenuation in helium II at temperatures below
1.5 K give values for $\alpha, \alpha'$ around 10$^{-2}$,
providing thus a negligible correction to the unperturbed
velocity $v_s$.

On the other hand, it is important to notice that according to
the general equation (4.1) derived in this work, the expectation
value of the free velocity operator (2.6), whose complex
counterpart reads
\begin{equation}
v = \frac{p}{M} + i \frac{\Omega}{2}\,z
\end{equation}
satisfies the evolution law
\begin{equation}
\langle v(t) \rangle= \left[\langle v(0) \rangle - v_s \left( 1-
\frac{\mu_0}{\beta}\right)\right] e^{\displaystyle i \Omega
\beta t} + v_s\left( 1 - \frac{\mu_ 0}{\beta}\right).
\label{mono}
\end{equation}
with $\mu_0 = \mu_{\Omega=0}$ (cf. Eq. (3.5)). Equation
(\ref{mono}) is remarkably close to the above expression
(\ref{feno}) for vanishing normal fluid velocity. Indeed, for a
normal fluid at rest, Eq.   ({\ref{mono}) is of the form
(\ref{feno}), with coefficients
$\tilde{\gamma}_0,\tilde{\gamma}'_0$ (or $\tilde{\alpha},
\tilde{\alpha}'$),  given by
\begin{eqnarray}
\Omega\,(\mu + i \gamma) &=& - (\tilde{\gamma}'_0 - i
\tilde{\gamma}_0)\,\frac{l}{M}
\nonumber
\\
\frac{\mu_0}{\beta}& =& \tilde{\alpha}'+i\,\tilde{\alpha}.
\label{mano}
\end{eqnarray}
Elimination of $\mu$ and $\gamma$ gives the relationship
\begin{eqnarray}
\tilde{\alpha}& =&   \frac{\displaystyle
\rho_s\,q\,h\,\tilde{\gamma}_0}{\displaystyle
\tilde{\gamma}_0^2+(\rho_s\,q\,h-\tilde{\gamma}'_0)^2}\,|\mu_0|
\nonumber
\\
\tilde{\alpha}'& =&
{\frac{\displaystyle
\rho_s\,q\,h\,( \tilde{\gamma}'_0-\rho_s\,q\,h)}{\displaystyle
\tilde{\gamma}_0^2+(\rho_s\,q\,h - \tilde{\gamma}'_0)^2}}\,|\mu_0|
\end{eqnarray}
The interesting similitude between these relations
and those in Eqs. (\ref{rela}) gives support to the
conjecture that the present model embodies  substantial aspects
of the mechanism responsible of damped vortex motion in
superfluid helium. The differences between the relationships
characterizing the phenomenological model in Eqs. (4.8), and the
present ones  in Eqs. (4.12),
are due to the fact that the structure of the drag force is not
identical in both approaches. In fact, a close look at Eqs.
(4.2), (4.3) and (4.6) shows that the Hamiltonian
description gives rise to an extra component of the force,
proportional to the relative two fluid velocity $v_n - v_s$.
This supplementary component is not removed by the assumption
that the normal fluid lies at rest; however, it is also
worthwhile to keep in mind that the assumption that the drag
force is proportional to ${\bf v} - {\bf v}_n$ applies under the
hypothesis of vanishing vortex mass \cite{Don1}. If one is
interested in  getting rid of the extra force, a
different  model should be selected, so as to bring the two
fluid dynamics into the picture. Such an improvement does not
consist of a simple modification of the Hamiltonian (2.5);
instead, a totally different formulation is required stemming
from a Hamiltonian description of the two fluids to which a
suitable copling is incorporated.  This philosophy fits more
specifically the spirit of macroscopic, fluiddynamical models
and is thus beyond the scope of the present  work.

As a final remark, we wish to recall that every time dependent
quantity here presented owes this dependence to the special model
feature that makes room to a finite, although unknown, inertial
coefficient of the vortex. This parameter rules the evolution
since it appears in both the conservative and the decay time
scale (cf. Eqs. (\ref{vl}), ({\ref{feno}) and ({\ref{mono})); we
then realize that as pointed out in Ref. \cite{dem2},
experimental detection of the time dependent regime would thus
provide a  means of measuring the vortex inertia. It should be
kept in mind that the present results concerning the dynamics
 cannot be extrapolated down to $M = 0$;
in fact, the free Hamiltonian (2.1), the frequency (2.3), the
vortex - reservoir Hamiltonian (2.5) and the velocity (2.6)
become meaningless in such a case. However, the asymptotic
velocity does not depend upon the mass, since its value 
causes the Magnus and the drift force to cancel each other in
the absence of inertial effects.

\section{Discussion and summary}

	Let us now examine further the characteristics of the model
here presented and its relationship to the phenomenological description
of dissipation. On the one hand, it is important to keep in mind
that the vortex mass is assumed to vanish in the
phenomenological two - fluid model of mutual friction, where the
 velocity arises from the balance between the Magnus and
the drag forces; it should be noticed that this regime is also
the time - asymptotic form of a Newton - like equation of motion if the
vortex mass is finite \cite{Don2}. The precise value of the
vortex inertial coefficient is thus not important in the limiting
regime, although it influences the dynamics at finite times 
through the frequency  $\Omega$ (cf. Eqs. (2.3) and (\ref{feno})),
which is the relevant parameter of the model. In this context, it
is important to keep in mind that the coupling to the thermal
excitations further renormalizes the vortex mass;  in fact,
inspection of the effective Hamiltonian (\ref{heff}) shows us that the
kinetic energy has been changed into $M (1 + \mu) {\bf v}^2/2$.

	On the other hand, the phenomenological theories introduce a
mutual friction whose drift and dissipative components are
proportional to the relative two - fluid velocity ${\bf v}_n - {\bf
v}_s$. The vortex velocity, either with respect to the superfluid or
to the normal one, is determined by the force balance when inertial
effects vanish; consequently, it depends upon the parameters of the
drag. Instead, the present model should be regarded as a description
in the reference frame of the normal fluid, {\it i.e.}, both $v_s$ and
$d \langle z \rangle/d t$ refer to the local velocity ${\bf v}_n$ of
the heat reservoir in the neighborhood of the vortex. The
coefficients that measure the drag effects thus depend upon the
strength of the  coupling to the thermal excitations and upon
their dynamical response; however
the fluid dynamics of the elementary excitations is not explicitly
contemplated.

	We believe that the model here presented covers most aspects
of the description of dissipative dynamics of a vortex line in helium
II and opens possibilities towards further improvements, among which,
a definite one is the introduction of the motion of elementary
excitations, to properly account for mutual friction in the sense of
phenomenological theories. With respect to
previous calculations of the drag coefficients carried, for
example, in Refs.  \cite{MS,SD}, our model, being quantal in
nature, is not subjected to either the low - temperature
limitations of a hydrodynamical description as pointed out in
\cite{MS}, or to uncertainties associated to a classical
approach to the roton - vortex collisions \cite{SD}. It may be
also mentioned that the model holds as well for vortex motion in
liquid $^3$He; quantum statistics only enters the
characterization of the excitations making the heat reservoir,
which would consist of the zero sound phonons of the fermion
liquid. No special differences with the present results would be
expected in that case,  except from the fact that the
larger core size of vortices in $^3$He could probably enlarge
 the inertia parameter $M$ with a subsequent decrease
in the oscillation frequency $\Omega$.

\acknowledgements 
 
 We are pleased to acknowledge stimulating conversations with Dr. Manuel
Barranco.  This work has been supported by grants PID 34520092
from Consejo Nacional de Investigaciones Cient\'{\i}ficas y
T\'ecnicas, Argentina, and EX100/95 from Universidad of Buenos Aires.

\end{document}